\documentclass[twocolumn,prl,aps]{revtex4}
\usepackage{psfig}
\newcommand{\be}{\begin{equation}}
\newcommand{\ee}{\end{equation}}
\newcommand{\bea}{\begin{eqnarray}}
\newcommand{\eea}{\end{eqnarray}}
\newcommand{\bs}{\mbox{\boldmath $s$}}
\newcommand{\bb}{\mbox{\boldmath $b$}}

\begin{document}
\title{Scanning the quark-gluon plasma with charmonium}
\author{
Boris Z.~Kopeliovich$^{a,b,c}$
Alberto Polleri$^d$ and 
J\"org H\"ufner$^{a,d}$}
\affiliation{
             $^a$ Max Planck Institut f\"ur Kernphysik, Postfach 103980,
                  D-69029 Heidelberg, Germany\\
             $^b$ Institut f\"ur Theoretische Physik der Universit\"at, 
 	          D-93040 Regensburg, Germany\\
             $^c$ Joint Institute for Nuclear Research, Dubna, 141980 
                  Moscow Region, Russia\\     
	     $^d$ Institut f\"ur Theoretische Physik der Universit\"at, 
                  D-69120 Heidelberg, Germany}

\date{\today}
\begin{abstract}
 We suggest the variation of charmonium production rate with Feynman
$x_F$ in heavy ion collisions as a novel and sensitive probe for the
properties of the matter created in such reactions.  In contrast to the
proton-nucleus case where nuclear suppression is weakest at small $x_F$,
final state interactions with the comoving matter create a minimum at
$x_F=0$, which is especially deep and narrow if a quark-gluon plasma is
formed. While a particularly strong effect is predicted at SPS, at the
higher RHIC energy it overlaps with the expected sharp variation with
$x_F$ of nuclear effects and needs comparison with proton-nucleus data.
If thermal enhancement of $J/\Psi$ production takes over at the energies
of RHIC and LHC, it will form an easily identified peak, rather than dip
in $x_F$ dependence.  We predict a steep dependence on centrality and
suggest that this new probe is complementary to the dependence on
transverse energy, and is more sensitive to a scenario of final state
interactions.
 \end{abstract}
\maketitle    

Although modification of the charmonium ($\Psi$) production rate due
to final state interactions (FSI) with the matter created in
relativistic heavy ion collisions may serve as a probe for its
properties \cite{ms}, it is still a challenging problem how to
disentangle the nuclear suppression at the early stage of a collision,
when the nuclei propagate through each other, from the late stage FSI,
when charmonium with a low speed travels through the comoving debris
of the nuclei. Since the former seems to be so far the main source of
suppression, uncertainties which exist in understanding the early
stage nuclear effects are transferred to the interpretation of FSI.

Even the very manifestation in the data of final state suppression is
still disputed. Although the dependence of charmonium suppression on
impact parameter ($E_T$ dependence)  cannot be reproduced by the
simplest model, where the effect is related only to absorption in
cold nuclear matter \cite{na50}, a deeper insight into the dynamics of
nuclear collisions discovers more sophisticated nuclear
effects \cite{hk,hkp1,qiu} which might nearly explain the observed
$E_T$ dependence. Currently, only the falling $E_T$ dependence of
suppression for most central collisions perhaps signals about the
presence of late stage interactions, since it is apparently related to
fluctuations of the density of produced matter
\cite{capella,blaizot,hkp2,ind}. However, even in this case it is
still difficult to discriminate between the models of comoving hadrons
\cite{capella} and phase transition to the quark gluon plasma
\cite{blaizot}.

In this letter we suggest a new sensitive probe for interaction of
charmonium with the produced matter and possibly to discriminate
between the models \cite{capella,blaizot}. This is the dependence on
Feynman $x_F$ of the ratio $R_{AB}(x_F)$ of charmonium production
rates in nucleus-nucleus and $pp$ collisions, which is expected to
expose a deep minimum at small $x_F$ due to absorption effects in the
final state. On the other hand, the early stage of interaction with
the nuclei leads to a maximum at small $x_F$. The shape of the minimum
turns out to be sensitive to the properties of the produced matter, it
is deeper and narrower if a phase transition to a deconfined
quark-gluon plasma takes place. On the contrary, if thermal
enhancement \cite{bs,thews,frank} of $J/\Psi$ production takes over 
in the energy range of RHIC and LHC, it
will create a peak in the $R_{AB}(x_F)$ at $x_F=0$ which should be 
clearly identified comparing with $pA$ data.

\begin{figure}[t]
\centerline{\psfig{figure=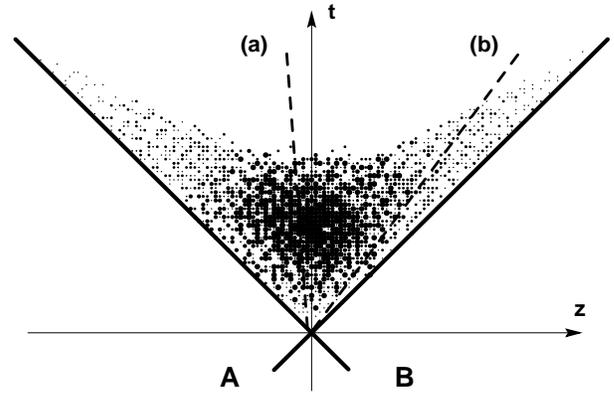,width=8cm,angle=-90}}
\protect\caption{Development of a nucleus-nucleus collision as function of
longitudinal coordinate $z$ and time $t$ in the c.m.  frame. Dashed lines
show trajectories of charmonia produced with small (a) and large (b)
values of $x_F$.} \label{cartoon} \end{figure}
 The production time of quarks and gluons $\sim 1/k_T$ correlates with
their transverse momenta and is usually longer that the one of a
charmonium $\sim 1/m_{\Psi}$. Therefore, interaction of the charmonium
with a comoving quark-gluon or hadronic gas factorizes from the early
stage of production and interaction with the nuclei. Depending on the
value of $x_F$, charmonium travels through different regions of the
produced matter, which is more dense at mid rapidity ($x_F=0$) while
it becomes more dilute with rising $x_F$. This is indicated in
Fig.~\ref{cartoon}, where different trajectories are shown by dashed
lines.  Thus, the charmonium scans the produced matter at different
rapidities depending on its value of $x_F$.  Apparently, the survival
probability of a charmonium propagating at $x_F=0$ through the most
dense medium is minimal, {\it i.e.} final state attenuation should
create a dip at $x_F=0$. The shape of this minimum depends on the
properties of the comoving matter and we give two model examples which
employ the ideas of either comoving hadrons \cite{capella} or phase
transition \cite{blaizot}.

Coming to the details of our calculation, we write  the differential
cross section for $\Psi$ production in nucleus-nucleus ($AB$)
collisions, as function of $x_F$ and impact parameter $\bb$, as  
\bea
\frac{d^3 \sigma_{AB}^{\Psi}}{d^2\bb\,dx_F} &\!\!=\!\!&  \frac{d
\sigma_{\mbox{\scriptsize $NN$}}^{\Psi}}{dx_F} \int \!d^2\bs \,\
T_A(\bs)\ T_B(\bb-\bs) \label{abexact} \\ & &\ \ \ \ \ \ \ \ \ \times\
S_{NUC}(\bb,\bs,x_F)\ S_{FSI}(\bb,\bs,x_F)\,,  \nonumber   
\eea  
where $T_A$ and $T_B$ are the nuclear thickness functions, while
$S_{NUC}$ and $S_{FSI}$ represent the modification factors due to
nuclear effects and FSI with the produced medium,
respectively.  

We now examine the features of nuclear suppression $S_{NUC}$ in $AB$ 
collisions, setting for the moment $S_{FSI} = 1$.
To perform calculations we employ the approximation of factorization
\be
S_{NUC}(\bb,\bs,x_F) =
S_{pA}(\bs,x_F)\,S_{pB}(\bb-\bs,-x_F)\ ,
\label{factorization}
\ee
which is accurate within the dynamics discussed below. The
observed $x_F$ dependence of the suppression factor $S_{pA}(\bs,x_F)$ 
at the energies of SPS \cite{na3,katsanevas} is well
understood in terms of absorption, energy loss \cite{kn,eloss} and
formation time effects \cite{kz91,hk-prl}. Therefore, we can use
model predictions for $S_{pA}(\bs,x_F)$. The agreement with the data is
very satisfactory and allows a reliable base-line calculation for the $AB$ 
case \cite{e-loss}.

\begin{figure}[htb]
\centerline{\psfig{figure=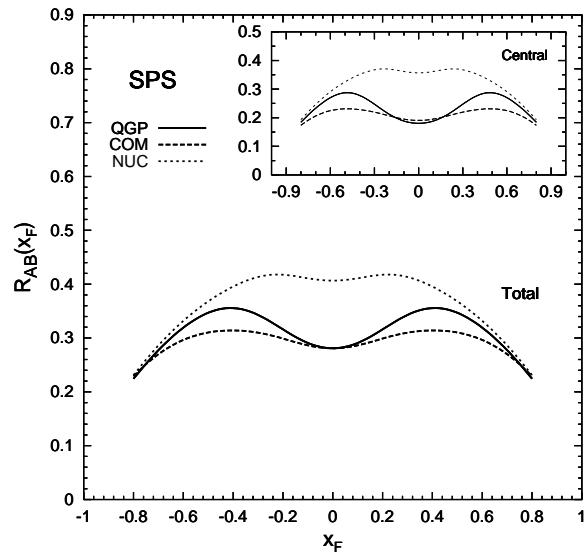,width=8cm}}
\protect\caption{Ratios of cross sections of charmonium production as 
function of $x_F$ for PbPb collisions at SPS, with respect to the $pp$ 
case. The dotted curve presents the effect of nuclear suppression alone.
Dashed and full lines refer to the full suppression
resulting from nuclear effects and also from interaction with plasma 
or comovers, respectively.  The
large and small panels represent the results for total cross section and
central collisions, respectively.}
\label{spsfig}
\end{figure}

Nuclear suppression as function of $x_F$ at given impact parameter $\bb$ of 
$AB$ collision is given by the ratio,
\be
R_{AB}(x_F,\bb) = \frac{1}{T_{AB}(\bb)}
\frac{d \sigma_{AB}^{\Psi}/d^2\bb\,dx_F(x_F,\bb)}
{d \sigma_{pp}^{\Psi}/dx_F(x_F)}\,.
\ee
Averaging with the thickness function $T_{AB}$, we also
compute the ratio of total cross section in $AB$ collisions with respect
to the $pp$ case, {\it i.e.}
 \be
R^{tot}_{AB}(x_F) = \frac{1}{AB} \int\!d^2\bb\ T_{AB}(\bb)\ 
R_{AB}(x_F,\bb)\,.
\ee
 These ratios for PbPb collisions at $E_{lab}=158$ GeV integrated over $\bb$
and at $\bb=0$ are plotted with dotted curves labeled ``{\footnotesize
$NUC$}'' in Fig.~\ref{spsfig} in large and small panels, respectively. One can
notice an approximately flat behavior for $|x_F| \leq 0.3$ and a rapid fall
off at large $x_F$.

\begin{figure}[htb]
\centerline{\psfig{figure=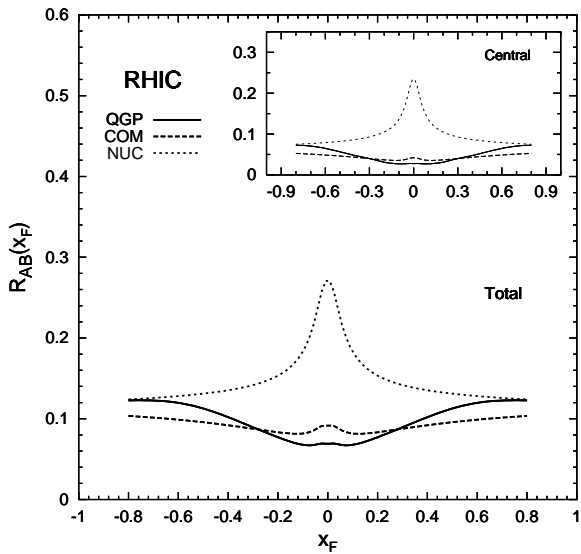,width=8cm}}
\protect\caption{The same as in Fig.~\ref{spsfig}, but for gold-gold 
collisions at RHIC energy $\sqrt{s}=200\,GeV$.}
\label{rhicfig}
\end{figure}

The mechanism of nuclear suppression completely changes at the high
energies of RHIC. Energy loss effects vanish, but coherence effects
come into play \cite{kth}. The conventional probabilistic treatment of
production and absorption of the $c \bar c$ pair becomes incorrect,
since amplitudes at different space-time points become coherent and
interfere. This leads to additional suppression which is an analog to
shadowing of $c$-quarks in DIS. Moreover, gluon shadowing becomes the
main source of suppression of $\Psi$ production.  While these
coherence effects barely exist at SPS, their onset has been already
observed at Fermilab \cite{e866} where they lead to a nuclear
suppression at large $x_F$ approximately as strong as at the SPS
energies, in spite of a substantial reduction of the energy loss
effects. Since the parameter free calculation of the coherence
effects performed in \cite{kth} are in a good accord with the Fermilab
data \cite{e866}, we can rely on it in order to predict the nuclear
suppression factor using eq.~(\ref{factorization}) for AuAu collision
at $\sqrt{s}=200$ GeV, shown by dotted curves in the two panels of
Fig.~\ref{rhicfig}.  One can see a new feature, absent in the SPS
case. The strong coherence effects, mostly gluon shadowing, form a
rather narrow peak in the $x_F$ dependence of nuclear suppression.

Unfortunately, predictions for $S_{FSI}$ in (\ref{abexact}) are still
ambiguous, whatever properties of the produced matter are assumed.
Nevertheless, to examine the scale of the expected effect in the $x_F$
distribution, we consider two popular scenarios of FSI. As we will
see, the shape of $x_F$ distribution turns out to be quite sensitive
to the distinction between the models.

The first model \cite{capella} assumes that the medium is composed by
comoving hadrons and provides the suppression factor 
 \be
S^{COM}_{FSI}(\bb,\bs) = 
\exp\!\left[- \sigma_{co}\,n_{co}(\bb,\bs)
\ln\left(\frac{n_{co}(\bb,\bs)}{n_{fo}}\right)\right], 
\label{fsicom}
 \ee 
 where $n_{co}(\bb,\bs)$ is the comover density in impact parameter
plane, the freeze-out density $n_{fo}=1.15$ fm$^{-2}$ and $\sigma_{co}
= 1$ mb are fitted parameters. In this model, the absorption cross
section controlling the factor $S_{NUC}(\bb,\bs,x_F)$ in
(\ref{abexact}) is adjusted at $\sigma_{\mbox{\scriptsize $\Psi
N$}}^{abs} = 4.5$ mb.

The scenario of ref.~\cite{blaizot} considers the possibility of QGP
formation and assumes that charmonium is not absorbed at all unless
the latter experiences a phase transition, which occurs when the
transverse density of participants $n_p(\bb,\bs)$ is close to the
critical value $n_{cr}$, yielding the suppression factor
 \be
S^{QGP}_{FSI}(\bb,\bs) = \frac{1 + \tanh\{\lambda\,[n_{cr} - 
n_p(\bb,\bs)]\}}{2}\,.
\label{fsiqgp}
 \ee
 Here $n_{cr} = 3.4$ fm$^{-2}$ and the smearing factor $\lambda = 2$
fm$^2$ are fitted parameters. In this model the nuclear absorption
cross section is adjusted to $\sigma_{\mbox{\scriptsize $\Psi
N$}}^{abs} = 6.4$ mb and is larger than in the comover model.

Both models are fitted to describe the observed $E_T$ dependence of
suppression at the mean value $\bar x_F=0.15$ of the NA50 experiment.
In order to incorporate an $x_F$ dependence into these models one must
admit that the density and properties of the produced matter which is
scanned by charmonium as is illustrated in Fig.~\ref{cartoon}, varies
with rapidity or $x_F$. $\Psi$s with small values of $x_F$ cross the
medium in its most dense part, while if their values of $x_F$ are
large, they interact only with a dilute matter. We assume that the
rapidity distributions of the produced matter traversed by $\Psi$ and
of the observed hadrons are proportional. Therefore, we re-scale the
densities of comovers $n_{co}$ in (\ref{fsicom}) and of participants
in (\ref{fsiqgp}) as
 \be
n(\bb,\bs) \ \Rightarrow\ n(x_F,\bb,\bs) 
= r(x_F)\ n(\bb,\bs)\,,
\label{100}
\ee
where the $x_F$-dependence is introduced by the factor
\be
r(x_F) = \frac{dN^h/dy}{dN^h_{SPS}/dy(x_F=0.15)}\,,
\label{200}
 \ee
 which is assumed to be independent of impact parameter of collision.
This approximation is supported by data from the SPS \cite{na49} and
RHIC \cite{phobos}. The rapidity $y$ of the comoving matter is the
same as that of charmonium. The latter is related to the $x_F$ of
$\Psi$ as $y(x_F) = (1/2)  \log(x_1/x_2)$. Here $x_{1,2} =
k_\Psi^{\pm}/k_N^{\pm}$ are the fractions of light-cone momenta of the
colliding nucleons carried by $\Psi$.  The normalization
$r(x_F=0.15)=1$ at SPS guarantees that the models under discussion are
unmodified for the kinematics of the NA50 experiment.

Note that the number of participants in eq.~(\ref{100}) is not meant to
depend on $x_F$. In fact, a phase transition in the plasma model should
be a function of the produced energy, which is assumed in
eq.~(\ref{fsiqgp})  to be proportional to the number of participants with
a coefficient dependent on $x_F$. This $x_F$ dependence is effectively
implemented in eq.~(\ref{100}).

We parametrize the rapidity density as the sum of two Gaussians
 \be 
\frac{dN^h}{dy}(y) = 
N\,\frac{\exp(- y_{-}^2/\Delta^2) + 
\exp(-y_{+}^2/\Delta^2)}{2\,\exp(- Y^2/\Delta^2)}\,, 
 \ee 
 symmetrically shifted from mid-rapidity by the amount $Y$, appearing
in $y_{\pm} = y \pm Y$, and characterized by a width $\Delta$. The
rapidity density is normalized to $N$ at mid-rapidity. The set of
parameters $N$, $Y$ and $\Delta$ is compared to SPS and RHIC
measurements of hadron spectra. For SPS we take the negative hadrons
rapidity distribution in central events \cite{na49} and are able to
reproduce it with $\Delta_{SPS} = 1.5$ and $Y_{SPS} = 0.75$. We scale
it up by a factor 3 to the total number of hadrons, which is about
$N_{SPS} = 600$ at mid-rapidity, considering that most of the hadrons
are pions. For RHIC we first take the charged hadrons pseudo-rapidity
($\eta$) distribution, measured at $\sqrt{s} = 130$ GeV \cite{phobos},
also for central events, and convert it to the one in rapidity with
the transformation 
 \be 
d/d\eta = \left[1 + \left(m/p_\perp
\cosh\eta\right)^2 
\right]^{^{\mbox{\scriptsize$-1/2$}}}\!\!\!\!
d/dy\,,
 \ee 
 assuming an average mass $m = 200$ MeV for the hadrons and a mean
value of the transverse momentum $p_\perp = 0.5$ GeV. We then scale it
up by a factor $3/2$ to get the total number of hadrons, and by a
factor $(200/130)^{0.36}$, to extrapolate to $\sqrt{s} = 200$ GeV. The
power energy dependence, $s^\delta$, of the inclusive cross section is
dictated by the AGK cutting rules \cite{agk}, and we fixed $\delta =
0.18$ interpolating between the measurements at 56 and 130 GeV.  This
is close to $\delta = 0.17$ found from data on inclusive pion
production in $pp$ collisions \cite{likhoded}. Neglecting for
simplicity that the width of the rapidity distribution also depends on
the collision energy, we obtain the values $\Delta_{RHIC} = 2.4$ and
$Y_{RHIC} = 1.75$, with a normalization $N_{RHIC} = 1000$.

The results for SPS are shown in Fig.~\ref{spsfig}.  The curves
correspond to nuclear effects only (dotted), model of comovers
(dashed) labeled ``{\footnotesize $COM$}'', and to the plasma model
(full) labeled ``{\footnotesize $QGP$}'', respectively.  The
``{\footnotesize $COM$}'' and ``{\footnotesize $QGP$}'' models include
nuclear suppression as it is calculated in \cite{capella,blaizot}.  
The big and small panels show $R^{tot}_{AB}(x_F)$ and
$R_{AB}(x_F,\bb=0)$, respectively. One clearly observes that the
plasma model leads to a more pronounced dip, while the comover model
produces only a slight depression. This feature is due to FSI with the
created medium and reflects its properties.

The predictions for RHIC are depicted in Fig.~\ref{rhicfig}. Nuclear
effects form a rather narrow peak (dotted curve) in $R_{AB}(x_F)$.
Applying calculated with \cite{capella,blaizot} FSI suppression, which
at RHIC is a very narrow dip which extends over about the same range
in $x_F$ as the peak caused by nuclear effects, one ends up with a
rather flat $x_F$ dependence. Note that the small irregularity around
of $x_F=0$ is an artifact of the used parameterizations and is within
the theoretical uncertainty. Since the FSI effect can be well singled
out comparing with $pA$ collisions, we emphasize the necessity of $pA$
measurements. Indeed, we do not expect a dramatic variation of the
shape with impact parameter since no peak is expected for peripheral
collisions.

Concluding, we suggest a novel probe for the matter created in heavy
ion collisions which is scanned by a charmonium produced with
different $x_F$. At SPS, we predict a pronounced minimum of survival
probability at $x_F=0$ which can occur uniquely by interaction of
$\Psi$ with the QGP. Its properties define the shape and depth of the
minimum, which varies with impact parameter, it is maximal for central
collisions and disappears for peripheral ones. In the comover case the
survival probability turns out to be rather flat. At the energies of
RHIC coherence effects form a narrow peak at $x_F = 0$, whose height
is substantially reduced due to the two different medium effects
considered. Although it will be difficult to distinguish the two FSI
scenarios, the prediction is rather interesting and can be confirmed
after comparing to $pA$ data taken at the same energy.

A possibility of FSI enhancement of $\Psi$, rather than suppression,
due to fusion of produced $c \bar c\,$s has been recently suggested
\cite{bs,thews,frank} to be present at RHIC. In this case FSI would
enhance the maximum produced by the nuclear factor $S_{NUC}$ in
(\ref{abexact}) leading to a dramatic effect easily observable.
However, while gluon shadowing leads to a very strong suppression of
direct $\Psi$s it should diminish even more (by square of that) the
fusion mechanism. This interesting problem needs further study.

Note that similar scanning is also possible with other hard processes,
for example jet quenching \cite{miklos}. One should measure di-jets or
di-hadrons in back-to-back geometry, {\it i.e.} with $\vec p_T^{\,1}
\!=\! -\vec p_T^{\,2}$ and $x_F^{1} \!=\! -x_F^{2}$. The energy loss
effect in a plasma is expected to have a similar $x_F$ dependence,
{\it i.e.} to provide stronger quenching at $x_F=0$.

\vspace{2mm}

\noindent {\bf Acknowledgments}:  Partial support by the grant from the
Gesellschaft f\"ur Schwerionenforschung Darmstadt (GSI), No.~GSI-OR-SCH,
and by the Federal Ministry BMBF grant No.~06 HD 954 is acknowledged.


\begin{thebibliography}{99}

\bibitem{ms} T.~Matsui and H.~Satz, 
Phys. Lett. {\bf B178} (1986) 416;

\bibitem{na50} M.C.~Abreu et al. (NA50 Coll.),
Phys. Lett. {\bf B477} (2000) 28. 

\bibitem{hk} J.~H\"ufner and B.Z.~Kopeliovich, Phys. Lett. {\bf B445}
(1998) 223; J.~H\"ufner, Y.B.~He and B.Z.~Kopeliovich, Eur. Phys. J.
{\bf A7} (2000) 239;

\bibitem{hkp1} J.~H\"ufner, B.Z.~Kopeliovich and A.~Polleri, {\sl
Excitation of Color Degrees of Freedom of Nuclear Matter and $J/\psi$
Suppression}, hep-ph/0010282, to appear in Eur. Phys. J. {\bf A};

\bibitem{qiu} J.~Qiu, {\sl $J/\Psi$ Production and Suppression in 
Nuclear Collisions}, talk at Quark Matter 2001, Stony Brook, January 
15-20, 2001.

\bibitem{capella}
N.~Armesto, A.~Capella and E.G.~Ferreiro,
Phys.\ Rev.\ C {\bf 59} (1999) 395;
A.~Capella, E.G.~Ferreiro, and A.B.~Kaidalov,
Phys. Rev. Lett. {\bf 85} (2000) 2080;

\bibitem{blaizot} 
J.P.~Blaizot and J.Y.~Ollitrault,
Phys. Rev. Lett. {\bf 77} (1996) 1703;
J.P.~Blaizot, P.M.~Dinh and J.Y.~Ollitrault,
Phys.\ Rev.\ Lett.\ {\bf 85} (2000) 4012;

\bibitem{hkp2} J.~H\"ufner, B.Z.~Kopeliovich and A.~Polleri, {\sl
Fluctuations of the transverse energy in Pb+Pb collisions and $J/\psi$
suppression}, nucl-th/0012003;

\bibitem{ind} A.K.~Chaudhuri, {\sl Transverse energy distributions and
$J/\psi$ production in Pb+Pb collisions}, hep-ph/0102038; 

\bibitem{bs} P.~Braun-Munzinger, J.~Stachel, Phys. Lett. {\bf B490}
(2000) 196;
 
\bibitem{thews} R.L.~Thews, M.~Schroedter, J.~Rafelski, J.
Phys. {\bf G27} (2001) 715.

\bibitem{frank}
M.I.~Gorenstein, A.P.~Kostyuk, H.~Stoecker, W.~Greiner,
``{\sl $J/\psi$ suppression and enhancement in Au+Au collisions at 
the BNL RHIC}'', hep-ph/0104071;

\bibitem{na3} J.~Badier {\it et al.} (NA3 Coll.), Z. Phys.
{\bf C20} (1983) 101;

\bibitem{katsanevas} The E537 Coll., S.~Katsanevas et al.,
Phys. Rev. Lett., {\bf 60} (1988) 2121;

\bibitem{kn} B.Z.~Kopeliovich and F.~Niedermayer,
``{\sl Nuclear screening in $J/\psi$ and lepton pair production}''
preprint JINR-E2-84-834, available at KEK library:
http://www-lib.kek.jp/cgi-bin/img\_index?8504113;

\bibitem{eloss} M.B.~Johnson et al., Phys. Rev. Lett. {\bf 86} (2001)  
4483

\bibitem{kz91} B.Z.~Kopeliovich and B.G.~Zakharov, Phys. Rev. {\bf D44} 
(1991) 3466;

\bibitem{hk-prl} J.~H\"ufner and B.Z.~Kopeliovich, Phys. Rev. Lett. {\bf 
76} (1996) 192

\bibitem{e-loss} J.~H\"ufner, B.Z.~Kopeliovich and A.~Polleri, 
{\sl Energy loss and charmonium production in heavy ion collisions}, 
paper in preparation;

\bibitem{kth} B.Z.~Kopeliovich, A.V.~Tarasov and J.~H\"ufner,
``{\sl Coherence Phenomena in Charmonium Production off Nuclei
at the Energies of RHIC and LHC}'', hep-ph/0104256;
B.Z.~Kopeliovich, {\sl Charmonium production off nuclei: from SPS to 
RHIC}, talk given at Quark Matter 2001, Stony Brook, January 15-20, 
2001, hep-ph/0104032;

\bibitem{e866} M.J.~Leitch {\it et al.} (E866 Coll.), 
Phys.\ Rev.\ Lett.\ {\bf 84} (2000) 3256;

\bibitem{na49} H.~Appelshauser {\it et al.} (NA49 Coll.),
Phys.\ Rev.\ Lett.\ {\bf 82} (1999) 2471;

\bibitem{phobos} G.~Roland {\it et al.} (PHOBOS Coll.), ``{\sl
Results from the PHOBOS experiment at RHIC}'' talk given at Quark Matter
2001, Stony Brook, January 15-20, 2001;

\bibitem{agk}
V.A.~Abramovsky, V.N.~Gribov and O.V.~Kancheli,
Yad.\ Fiz.\ {\bf 18} (1973) 595;

\bibitem{likhoded} A.~Likhoded et al., Int. J. Mod. Phus. {\bf A6} (1991) 913;

\bibitem{miklos} M.~Gyulassy and M.~Pl\"umer, Nucl. Phys. {\bf B346}
(1989) 1.

\end{thebibliography}
\end{document}